\documentstyle[11pt,aaspp4,epsf]{article}
\begin{document}

\newcommand{\msun}{\mbox{${\cal M}_\odot$}}
\newcommand{\lsun}{\mbox{${\cal L}_\odot$}}
\newcommand{\kms}{\mbox{km s$^{-1}$}}
\newcommand{\HI}{\mbox{H\,{\sc i}}}
\newcommand{\mhi}{\mbox{${\cal M}_{HI}$}}
\def\hst{{\it HST}}
\def\rsun{{\rm\,R_\odot}}
\def\etal{{\it et al.}}
\newcommand{\HII}{\mbox{H\,{\sc ii}}}
\newcommand{\NII}{\mbox{N\,{\sc ii}}}
\newcommand{\SII}{\mbox{S\,{\sc ii}}}
\newcommand{\nan}{Nan\c{c}ay}
\newcommand{\darkV}{$\frac{{\cal M}_{HI}}{L_{V}}$}
\newcommand{\dark}{$\frac{{\cal M}_{HI}}{L_{B}}$}
\newcommand{\am}[2]{$#1'\,\hspace{-1.7mm}.\hspace{.0mm}#2$}
\newcommand{\as}[2]{$#1''\,\hspace{-1.7mm}.\hspace{.0mm}#2$}
\def\lsim{~\rlap{$<$}{\lower 1.0ex\hbox{$\sim$}}}
\def\gsim{~\rlap{$>$}{\lower 1.0ex\hbox{$\sim$}}}
\newcommand{\iso}{sech$^{2}(z)$}


\title{The Extraordinary `Superthin' Spiral Galaxy UGC~7321. II. 
The Vertical Disk Structure}
\vskip0.5cm
\author{L. D. Matthews\altaffilmark{1}}

\altaffiltext{1}{National Radio Astronomy Observatory, 520 Edgemont Road, 
Charlottesville, VA 22903 USA, 
Electronic mail: lmatthew@nrao.edu}

\singlespace
\tighten
\begin{abstract}
We explore the vertical light distribution  as a function of
galactocentric radius in the edge-on ($i$=88$^{\circ}$)
Sd ``superthin'' galaxy UGC~7321. UGC~7321 is a low-luminosity
spiral ($M_{B,i}=-17.0$) 
with a diffuse, low surface brightness 
stellar disk and no discernible bulge component.
Within $\sim$\am{0}{25} ($\sim$725~pc) of 
the disk center the global luminosity profile of UGC~7321 can
be reasonably characterized
by an exponential function with a scale height $h_{z}\sim$\as{2}{9} 
($\sim$140~pc) in $H$ and $h_{z}\sim$\as{3}{1} ($\sim$150~pc) in $R$,
 making this among the thinnest
galaxy disks know. Near the disk center we derive a ratio of disk scale length
to global disk scale height $h_{r}/h_{z}\sim$14 in both $H$ and $R$; near the
edge of the disk, $h_{r}/h_{z}\sim$10. At  intermediate 
galactocentric radii (\am{0}{25}$\le|r|\le$\am{1}{5}), the disk of UGC~7321
becomes less peaked
than an exponential near the galactic
plane. At these radii the vertical
luminosity profiles can be well reproduced 
by a linear combination of two isothermal
disk components of differing scale heights. 
These fits, together with the strong disk color gradients 
by Matthews, Gallagher, \& van Driel (1999), suggest
that UGC~7321 has multiple disk
subcomponents comprised of stellar populations with different ages and
velocity dispersions. Thus even examples of
the thinnest pure disk galaxies exhibit
complex structure and signatures of dynamical
heating. 
\end{abstract}


\section{Introduction}
\subsection{Background}
Studies of the vertical light distributions of galaxy disks can offer vital
clues toward understanding the physics and evolutionary histories
of these systems.  For
example, vertical scale height measurements can be related to the
stellar velocity dispersions and density profiles of disks, so
constraints on their stability and dynamical evolution can be derived
(e.g., Fuchs \& Wielen 1987; Dove \& Thronson 1993; Bottema
1993,1997; van der Kruit 1999a).  When scale height measurements are 
combined with vertical
color profiles, information can be gleaned on star formation histories
and dynamical heating mechanisms (e.g., Just \etal\ 1996).

Van der Kruit \& Searle (1981a,b;1982a,b; see also Spitzer 1942) 
proposed that galaxy disks
could be modelled as self-gravitating locally isothermal sheets whose vertical
light profiles $L(z)$ are described by
\begin{equation}
L(z)=L_{o}{\rm sech}^{2}(z/z_{0})
\end{equation}
\noindent where $L_{o}$ is the luminosity surface density, $z$ is the
vertical distance from
the midplane of the galaxy, and $z_{0}$ is the  scale height. In this
model, the vertical velocity dispersion $\sigma_{z}$, is
independent of $z$ at a given galactocentric radius $r$.

Van der Kruit (1988) later argued that disks may deviate from
isothermal near the galaxy plane, hence a better description of their
overall vertical light profiles might be given as
\begin{equation}
L(z)=L_{o}{\rm sech}(z/h_{z})
\end{equation}
\noindent which differs from a sech$^{2}(z)$ function only at small
$z$. Here,  $h_{z}=z_{0}/2$. This model has been applied successfully to
fit the light distributions of some nearby galaxies (e.g., Barnaby \&
Thronson 1992; Rice \etal\ 1996).

As new and better data have become available in recent years,
measurements of the vertical
light distribution of our own Galaxy, as well as a number of external galaxies,
have often pointed in favor of an exponential $z$ light
distribution of the form
\begin{equation}
L(z)=L_{o}{\rm exp}(z/h_{z})
\end{equation}
\noindent (e.g., Gilmore \& Reid 1983; Pritchet 1983; Wainscoat,
Freeman, \& Hyland 1989; Aoki \etal\ 1991;
van Dokkum \etal\ 1994; Barteldrees \& Dettmar 1994; 
Kodaira \& Yamashita 1996; de Grijs \& Peletier 1997; 
Fry \etal\ 1999). Finally, de Grijs,
Peletier, \& van der Kruit (1997)  have explored a more general family
of fitting functions, which include relations 1-3 as special cases.

Part of the problem in distinguishing between these various relations
empirically is that all of these functions look similar at large $z$
values, while at small $z$ the data are often contaminated by dust. 
There has also been little agreement on which of these relations
should be favored on theoretical grounds. For example,
it has remained poorly determined how each of these various
distributions might be achieved physically,
how the vertical structures of galaxy disks should be expected to
change over time due to both internal and external processes, and to 
what degree the vertical structures of disks
are determined by the initial conditions of formation (cf. Burkert \&
Yoshii 1996) versus subsequent evolution (e.g.,  Fuchs \& Wielen 1987;
Just \etal\ 1996).  One way to gain insight into these issues is through
detailed investigations of the vertical light distributions of galaxies
spanning a range in properties. 

To date, most detailed studies of the vertical structures of
galaxies have
concentrated on moderately luminous galaxies with normal
surface brightness disks (e.g., de Grijs 
1997 and references therein). Very 
few analyses (e.g., Kodaira \& Yamashita 1996;
Fry \etal\ 1999) have probed the vertical disk
structures of the very latest spiral types, 
especially those
with low luminosities and
low optical surface brightness disks.
However, 
the vertical structures of such ``extreme late-type spirals'' are of particular
interest. For example, certain 
extreme late-type spirals
appear to represent some of the
least evolved nearby galaxy disks (e.g., Matthews \& Gallagher
1997). Comparisons of their vertical structures with theoretical
predictions can thus help to constrain
evolutionary scenarios for these galaxies and allow us to 
determine
how disk structure, dust contents, and
dynamical evolution may differ between galaxy disks of high and
low surface brightness.  Because many 
extreme late-type and low
surface brightness galaxies appear to be highly dark
matter-dominated systems (e.g., Bothun, Impey, \& McGaugh 1997 
and references therein), 
vertical structure measurements
of these systems combined with rotation curve and stellar velocity
dispersion measurements may ultimately hold the key
to unravelling the nature and distribution of dark matter in disk
galaxies (e.g., Zasov, Makarov, \& Mikhailova 1991; Bottema 1993,1997;
Matthews 1998; Swaters 1999). 

\subsection{The Vertical Structure of a `Superthin' Spiral}
In a previous paper (Matthews, Gallagher, \& van Driel 1999; hereafter
Paper~I), new multiwavelength data, including
optical and near-infrared (NIR) imaging data, were
presented for
the edge-on ``superthin'' spiral UGC~7321. Here we utilize
those datasets to probe
the vertical luminosity distribution and vertical scale height of 
this galaxy over a range of galactocentric radii. 

Visual
inspection alone reveals some of UGC~7321's
most interesting traits: a diffuse stellar
disk, large disk axial ratio, and complete absence of a 
discernible bulge component 
(Fig.~1). Some additional properties of UGC~7321 are summarized in
Table~1. The distance quoted in Table~1 ($10^{+3}_{-3}$~Mpc) is estimated from
resolved luminous stars in imaging data obtained with the {\it Hubble Space
Telescope} (Gallagher \etal, in preparation). 
This distance is twice that estimated by Tully, Shaya, \&
Pierce (1992) based on galaxy 
group assignments and peculiar motion analyses, indicating that UGC~7321
appears to be a rather isolated system, and is unlikely to have undergone any
recent perturbations.  

In Paper~I it was  shown 
that UGC~7321 has a low optical surface  brightness disk that exhibits
significant radial and vertical color gradients, and that it
suffers relatively minimal
internal extinction due to dust.  Together these trends were
interpreted as evidence that
UGC~7321 is highly ``underevolved'' disk in both a dynamical and in a
star-formation sense. Therefore it is of tremendous interest to compare
the vertical structure of this seemingly ``primitive'' galaxy
with that of more ``normal'' spirals.
Fortunately the  nearly edge-on geometry of UGC~7321  ($i\approx88^{\circ}$)
makes it ideally suited for such an analysis.

\section{The Data}

Optical and NIR imaging
observations are complementary for
the analysis of the vertical density structures of galaxies. NIR images
have the advantage that they are considerably less affected by dust 
than optical data, so the effects of
internal extinction on the observed brightness distribution are
minimized and the vertical light distribution can be probed even at
very small $z$-heights.
In addition, the light contribution from  recent star
formation is lower in the NIR, so NIR light profiles of
galaxies should give the best
measure of the radial and vertical distribution of the underlying
older stellar populations. Unfortunately, due
to the high NIR sky background,  NIR
observations are generally of lower signal-to-noise than 
CCD observations at optical wavelengths,
particularly for very low surface brightness galaxies. It
is therefore more difficult to trace the
light profiles of galaxies in the faintest outer disk regions  or
at large $z$ heights in the NIR. However,
 optical observations can be used to supply
supplemental information in these regimes, since dust extinction is 
usually minimal in those regions. Moreover, comparisons of galaxy
scale heights and scale lengths measured in multiple wavebands can
supply important information about disk evolution and star
formation histories. For these reasons, we undertake the analysis
of the vertical structure of the UGC~7321 disk utilizing both $H$-band
and $R$-band imaging observations.
Details regarding the acquisition and
reduction of these data can be found in Paper~I. 

\section{Radial Scale Length Measurements}
In Paper~I, a radial disk scale length for UGC~7321 was reported
for the $B$-band only. Here we rederive this measurement
using an alternate technique. We also 
measure scale lengths in the $R$ and $H$
bands for subsequent comparison with our vertical structural measurements at
these wavelengths.

As is well known, accurately measuring scale lengths for galaxy disks
can be a difficult task, and the numbers derived can vary greatly
depending on the technique used and the limiting isophote of the data
(e.g., Knapen \& van der Kruit 1991).
In Paper~I, a $B$-band radial scale
length of $44''\pm 2''$ was derived for UGC~7321 using isophotal
fits. This technique has the advantage of maximizing signal-to-noise
for obtaining a global disk scale measurement, although as 
was noted in Paper~I, the validity and accuracy of this technique begins
to break
down for galaxies viewed very near edge-on. The result can be
a tendency to slightly underestimate the true disk scale length in the
nearly edge-on case (see also de Grijs 1998).

An alternate method for measuring the radial scale lengths of
edge-on galaxies is through fits parallel to their major axis light
profiles. We adopt this technique in the present work.

Our data were first processed as described in Sect.~4.1 below. 
We then extracted light profiles along the
major axis from the $B$-, $R$-, and $H$-band data. The extracted
profiles were 12-pixels thick in $B$ and $R$ and 3 pixels thick in
$H$. Often a light profile is extracted slightly offset
from the true major axis in order to avoid the effects of
dust (e.g., Wainscoat, Freeman, \& Hyland 1989; Barteldrees \& Dettmar
1994; de Grijs 1998). 
Unfortunately, because the UGC~7321 disk is so thin, this is not
possible in the present case. To increase
signal-to-noise,
 we folded each of the radial profiles about $r$=0, and in the
case of the $R$- and $B$-band data, we have also smoothed the profiles
by a factor of 15 along the radial direction. The resulting surface
brightness profiles are shown in Fig.~2.  As our $H$-band data
are not photometrically calibrated, throughout this paper we illustrate
these data using an absolute
surface brightness scale estimated from a comparison with the
photometrically-calibrated Two Micron All Sky Survey (2MASS) $H$-band
data for UGC~7321\footnote{The Two Micron All Sky Survey photometric
parameters  were
accessed through http://sirtf.jpl.nasa.gov/2mass/.}.

To determine the scale lengths, we fitted each of the profiles in Fig.~2
using Eq.~5-9 of van der Kruit \& Searle (1981a; see also
Paper~I). The best fits were determined by eye and 
are overplotted on Fig.~2 as dashed lines. Our
derived measurements are as follows: $h_{r,H}=41''\pm3''$,
$h_{r,R}=43''\pm4''$, and $h_{r,B}=51''\pm4''$. The differences as a
function of wavelength reflect the radial color gradients present in the disk
of UGC~7321 (see Sect.~6.1.1; see also Paper~I).

From Fig.~2 it can be seen that a pure exponential provides only an
approximate fit to the data. Possible implications
of this are discussed
further in Paper~I. Because the disk of UGC~7321 is rather
diffuse, even our smoothed, folded  profiles are not 
very smooth at optical wavelengths. In addition, a 
light excess (discussed further in
Paper~I) is seen at small radii,
particularly in the $H$ band, while at large radii ($r\approx150''$)
it appears the disk may be truncated. Finally, over the interval
$70'' \le r \le 130''$, Fig.~2 shows there is also a light
excess visible 
in both the $B$ and $R$ bands.

\section{Vertical Structure Measurements}
\subsection{Extraction of the Vertical Light Profiles}
We explored the vertical structure of the UGC~7321 disk by examining
one-dimensional cuts extracted from our $H$- and $R$-band images
at a variety of galactocentric radii.
Before extracting the light profiles,
the images were sky-subtracted and the
IRAF\footnote{IRAF is distributed by the National Optical Astronomy
Observatories, which is operated by the Associated Universities for
Research in Astronomy, Inc. under cooperative agreement with the
National Science Foundation.} task ``imedit'' was used to  remove
cosmic rays, field stars, and 
background objects in the vicinity of the galaxy. Using field stars as
a reference, we found
the peak central brightness of UGC~7321 to occur in identical
locations in both the $H$ and the $R$ frames, and defined this to be
the galaxy center ($r$=0).

Beginning at $r$=0, we extracted 3- and 15-pixel ($\sim3''$) wide
cuts parallel to the galaxy minor axis from the $H$ and $R$ images,
respectively, at \am{0}{25}-\am{0}{5} intervals along the galaxy.
A position angle of 172$^{\circ}$ for the galaxy minor axis was adopted.
The locations of the extracted cuts are
illustrated in Fig.~1. Throughout this paper 
we use $z$ to denote the distance above or below
the galaxy midplane, and $r$ to denote the distance from the disk center.
For clarity we adopt the convention that $r$ is positive on the 
eastern (E) side
of the disk, and negative on the west (W) side, rather than
the usual cylindrical coordinates.

The widths of the extracted strips were selected
to obtain reasonable signal-to-noise without sacrificing
the ability to discern changes in scale height or disk structure as a
function of radius. Because in $H$-band,
our pixel size (\as{1}{09}) is non-negligible compared
with the thickness of the UGC~7321 disk, no binning was applied to
the cuts along the vertical direction; pixels in the optical data are smaller
(\as{0}{195}), but still no vertical binning was applied due to the very steep
slope of the light profiles.

\subsection{Gauging the Effects of Dust}
An important consideration in analyzing the vertical light structures
of galaxies is accounting for the effects of internal dust absorption
on the observed light distribution. Our optical data (Paper~I), as well
as $R$ and $I$ 
images of UGC~7321 obtained with the {\it Hubble Space Telescope} 
(Gallagher \etal, in preparation) reveal
that UGC~7321 clearly contains dust, although this galaxy lacks the
quintessential dust lane found in most brighter spirals 
viewed edge-on (cf. Howk \& Savage 1999). The dust distribution in UGC~7321 
instead appears   patchy and
clumpy, and cannot be well approximated by a uniform layer (see also
Matthews \& Wood, in preparation).
 
Fortunately for our purposes, the bulk of the
dust in UGC~7321 appears to be
confined to a region along the galaxy midplane only a few arcseconds thick
($-3''$\lsim$z$\lsim$7''$). This can also be 
seen in our
$R$-band vertical cuts (e.g., Fig.~8-10, discussed below), where 
indentations due to dust are visible. The slight asymmetry of these features
about $z$=0 confirms that UGC~7321 is not viewed exactly edge-on, but rather at
an inclination $i\approx$88$^{\circ}$ (see also Paper~I). 

In $H$-band, the effects of dust appear to be essentially 
negligible in UGC~7321. None of our $H$-band profiles  exhibit
obvious indentations due to dust (see Fig.~3-7, discussed below) and the
$H$-band contours show no evidence for the type of asymmetries or 
irregularities
expected from dust absorption (see Fig.~2 of Paper~I). 
For these reasons, and because of the patchy distribution of
dust in UGC~7321, in the present work
 no effort was made to model its effects on the light
profiles in a sophisticated manner.
Effects of
dust were therefore ignored in the $H$-band, while in the
$R$-band,  dust at small $z$ was compensated for by a simple extrapolation of
the profiles from intermediate $z$ regions (see also Kodaira \&
Yamashita 1996). 
The dust distribution of UGC~7321
and its detailed effects on the observed properties of
UGC~7321 have recently been investigated by Matthews \&
Wood (in preparation) using three-dimensional Monte Carlo radiation
transfer models. These sophisticated models confirm that dust
effects in $H$-band are quite small and that the simple dust
treatment we adopt here is adequate for our present purposes.

\subsection{The Fitting Technique}
Using a nonlinear least squares fitting technique based on the CURFIT
program of Bevington (1969), we attempted fits to the 
extracted vertical luminosity profiles  
with each of the three functions given by Eq.~1-3
(hereafter the `\iso', `sech$(z)$', and `exponential' functions).
These respective functions were convolved with a Moffat function (Moffat
1969) before fitting in order to mimic the effects of seeing on the
model light profiles.  The Moffat function is given as
\begin{equation}
p(r)=\frac{\beta - 1}{\pi\alpha^{2}}\left[1 +
\left(\frac{r}{\alpha}\right)^{2}\right]^{-\beta}
\end{equation}
\noindent with
\begin{equation}
{\rm FWHM} = 2\alpha\sqrt{2^{1/\beta} - 1},
\end{equation}
\noindent where $\alpha$ was solved for in each case.
We assumed  $\beta$=2.5 (Saglia \etal\ 1993), and the FWHM
for a point source in each band 
was determined using the mean value from fits to several
stars in each field.

Each point in the light profiles was weighted during fitting by a term
that accounted for Poisson noise and sky and flatfield uncertainties.
Initially, the profiles on either side of the midplane
[hereafter the $+z$ (north) and $-z$ (south) sides] were fitted
independently. This permitted us to search for systematic difference in
shape or scale height of the two sides of the profile
and provided a means of estimating the uncertainty in the fits (see below).

For each fit it was necessary to supply initial guesses for the
parameters $L_{o}$ and $z_{0}$ (or $h_{z}$). The fitting program then
used a chi-squared minimization technique to find an acceptable
fit. Refined values of $L_{o}$ and $z_{0}$ (or $h_{z}$) were then
inputted, and the process was repeated for several iterations before
converging at a final solution. All fits were also compared with
the data by eye as a second check of fit optimization.
To ensure proper handling of errors, the fitting was 
initially performed in linear
rather than magnitude
space.

\subsection{Fitting Round 1: The $H$-Band Data and One-Component Models}
To minimize any impact from dust and small-scale star-forming complexes,
we performed a first  round of fits  on the $H$-band
profiles.  Our $H$-band data have sufficient signal-to-noise and
field-of-view to
permit fits to the vertical profiles over the disk region 
$-$\am{1}{5}$\le r \le +$\am{2}{0}. 
Each individual profile was fitted over the interval
$-20''\le z \le +20''$, which was found by trial and error to optimize the
quality and stability of the fits.
From the initial set of fits to the $H$-band data, several interesting
features of
the vertical light distribution of UGC~7321 became apparent.

Fig.~3 illustrates the best fit 
along the disk minor axis ($r$=0). At this location,
the $H$-band vertical 
light profile is too strongly peaked at small $z$ values to be
reproduced by either a sech$(z)$ or \iso\ function. However, the  
profile can be reasonably well reproduced  over the full range of $z$
by a single exponential
function with a scale height $h_{+z,c}$=\as{3}{0} on the $+z$  side
of the galaxy and $h_{-z,c}$=\as{2}{74} on the $-z$  side. At the
adopted distance of UGC~7321 this corresponds to a mean scale
height of $\overline{h}_{z,c}\approx$140~pc. {\it This is one of the
smallest global
vertical scale heights ever reported for a galaxy
disk.}\footnote{Using NIR data, Kodaira \& Yamashita (1996) measure a 
global scale
height of $h_{z}=$158~pc for the pure-disk,
edge-on galaxy NGC~4244, making this one of the few reported values comparable
to that which we measure for UGC~7321. However, 
Fry \etal\ (1999) report a larger global exponential scale height for
NGC~4244 in the $R$
band ($h_{z}$=246~pc).} For example,
this is less than half  the global exponential scale height of the Milky Way
disk ($h_{z}\approx$300~pc; e.g., Freeman 1991).

Moving slightly away from the disk center, 
at $r=-15''$ we find the $H$-band light profile is still 
well fit by an exponential, but with a slightly
larger mean scale height ($\overline{h}_{z,-15''}$=\as{3}{4})
than measured at $r$=0 (Fig.~4a).
However, on the opposite side of the disk, at $r=+15''$, an exponential
with $\overline{h}_{z,+15''}=$\as{3}{63} fits the data nicely 
at large $z$, but at
small $z$ this function 
is seen to be somewhat too strongly peaked to reproduce 
the data (Fig.~4b). On the other hand, a 
sech$(z)$ function with $\overline{h}_{z}$=\as{2}{45} 
is slightly too rounded  at
small $z$, and leaves residuals in the profile wings.
Because of the exponential nature of the profiles at $r$=0 and $r=-15''$,
this deviation from an exponential cannot easily be attributed to inclination,
seeing, or resolution effects.
Moreover, we see this trend become more pronounced 
as we move to larger galactocentric radii. 

At $r=\pm$\am{0}{5} we again find that
the light profile can be fit by an exponential at large $z$, but  at
small $z$, the exponential is clearly
too strongly peaked to fit the data (Fig.~5). Instead, 
at small $z$ values the shape of the light profile
is
better reproduced by either
a sech$(z)$ or a \iso\ function, with the sech$(z)$ giving the best
overall fits. 

At larger galactocentric radii, the sech$(z)$ function continues to
provide the best global fits to the $H$-band data over the full range of $z$.
 The
results of these fits are shown in Fig.~6.  Table~2 summarizes the
mean derived scale heights $h_{z}$
from the one-component fits at various galactocentric radii.

The numbers
in Table~2 seem to suggest that the scale height of the UGC~7321 disk
increases with radius by over 30\% in the interval $-$\am{1}{5}$\le
r\le-$\am{0}{5}, and by over 40\% in the interval +\am{0}{5}$\le
r\le$+\am{2}{0}.  As discussed in Sect.~5, we estimate the errors in
our single-component model fits to be $\sim\pm$10\%, hence this
increase is statistically significant. 

In general, disk scale
heights of late-type spiral galaxies 
are found to increase by no more than a few per cent with
radius (e.g., de Grijs \& Peletier 1997), 
hence the large increase we measure for UGC~7321 is
surprising. One possibility is that there is significant flaring in
the outer disk due to decreasing self-gravity (see also
Capaccioli, Vietri, \& Held 1988). Another possibility is that the
one-component sech$(z)$ model fits are not adequate to describe the
data at all $z$ heights due to the presence of an additional disk
component. 
While the first possibility likely makes some
contribution to the trends we observe (see also Sect.~6.1.2), 
as we show below, it indeed
also appears
that the UGC~7321 structure is more complex than indicated by a
simple, one-component model.

\subsection{Fitting Round 2: The H-band Data and Two-Component
Models}

In Fig~5\& 6, it is clear that while the sech$(z)$ fits can be
argued to provide a reasonable,
rough global characterization of
the light profile shapes, they fall short of completely reproducing
the observed light profiles to within the observational errors over
the full range of $z$. In
particular, in the $r=\pm$\am{0}{5} and $r=\pm$\am{1}{0} profiles, where
signal-to-noise is high, small light excesses are visible in the data
at $|z|$\gsim$9''$, especially on the $+z$ side of the disk. This
excesses suggest that a second fit component may be necessary 
to fully reproduce
the observed vertical light distribution of UGC~7321.

To test this possibility, in 
Fig.~7a \& b we show revised fits to the $H$-band light profiles of UGC~7321
at $r=\pm$\am{0}{5},
this time using a linear combination of {\it two \iso}
components of
differing scale heights\footnote{Linear combinations of 
two sech$(z)$ components
also provide similarly good fits to the data (see Sect.~5).}. 
To derive the fits shown, we
initially fitted the $+z$ and $-z$ halves of each profile
independently. By using a  mean of these results, we arrived at a simple,
two-component \iso\ model that
provides an excellent and improved fit to the data over  the full range of
$z$. In these fits, the scale height
of the inner fit
component was taken to be $z_{0,1}$=\as{3}{8} (185~pc) 
 and the scale-height of the outer component was
taken to be $z_{0,2}$=\as{8}{7} (423~pc). The relative
light contributions of the two components were the same on both
the E and W 
sides of the disk. Only a  small difference in scaling of the
total profile was necessary to account for the slightly different peak
brightnesses on the two sides.

Even though we have used mean fit values in Fig~7, the two-component
fits still show a
clear improvement over the single-component sech$(z)$ models shown in
Fig.~6. Both the inner points and the profile wings are now reproduced to
within their error bars over the interval $-15''\le z\le+15''$.

To test the statistical significance of the improvement produced by
the extra light component in these new
models, we have applied a statistical $F$ test to compare the two fits 
(e.g., Bevington 1969; Chatterjee
\& Price 1977). 
Following Chatterjee \& Price (1977), the $F$ ratio applicable to
determining the significance of additional fit terms is defined as 
\begin{equation}
F=\frac{[\chi^{2}(1)-\chi^{2}(2)]/(p+1-k)}{\chi^{2}(2)/(n-p-1)}
\end{equation}

\noindent where $\chi^{2}(1)$ characterizes 
the ``full model''  with $(p+1)$ free parameters [here the two-\iso\
model], 
 $\chi^{2}(2)$ characterizes the ``reduced model''   with
$k$ distinct parameters [here the sech$(z)$ model], and $n$ is the
number of data points fitted. 

For this test, we presently limit ourselves to the interval
$-15''\le z\le+15''$. At $r=-$\am{0}{5}, for the 29 fitted data points,
we derive $\chi^{2}(1)$=42.94
and $\chi^{2}(2)$=11.79. At $r=+$\am{0}{5}, for the 25 fitted
data points, $\chi^{2}(1)$=18.72 and $\chi^{2}(2)$=7.34.
From the $F$ distribution probability table of Abramowitz \& Stegun
(1965; Table~26.9), we infer that in both instances there is a $>99.9$\%
probability that these
second fit components are warranted. From this we  deduce that  {\it at 
moderate galactocentric radii, the disk of UGC~7321 is not strictly 
isothermal or sech$(z)$-like 
over its full z-extent, but rather has a more complex structure}.

Over the full interval $-$\am{1}{5}$<r\le +$\am{1}{5}, 
we continue to find that 
the two-component model comprised of two \iso\ functions of scale
heights  $z_{0,1}=$\as{3}{8} and $z_{0,2}=$\as{8}{7}, respectively,
 produces superior fits to the data compared with the single-component
sech$(z)$ fits, based on both the $F$ statistic 
(at the $>90$\% certainty level) and visual inspection, provided that 
the relative
contribution to the total fit of the larger $z$-height component is
allowed to increase with increasing galactocentric
radius. A slightly more significant increase in the
contribution of the higher $z$-height component was required on the W side
of the galaxy than on the E side (see Fig.~7). 
We discuss possible implications of this
further below. Over this interval we find no evidence for scale height
changes in either component to within errors, although we cannot rule
out changes of \lsim10\%. 
Only at $r=+$\am{1}{5} and $r=+$\am{2}{0} do the
sech$(z)$ models and the two-component \iso\ 
models become formally indistinguishable in the $H$-band (see also
Sect.~4.6\&6.1.1).

\subsection{Fitting Round 3: R-Band Fits}
As was shown above, over the interval 15$''$\lsim$|r|$\lsim\am{1}{5},
the $H$-band light profiles perpendicular to the UGC~7321 disk are
best reproduced over their full $z$ extent by a model comprising 
a sum of at least two light
components.
Before interpreting this result further, 
we need to first explore the possibility that the
requirement for 
a second light component is simply an observational
artifact due to scattered light or
flatfielding problems in the $H$-band data.
Since the signal-to-noise of the $R$-band data is superior to that of
the $H$-band data at large $z$ values and at large $r$, 
these data can be used to provide a check on the results obtained
at $H$-band.

We began by attempting to fit the vertical $R$-band profile at $r$=0 in the
manner described above, successively
using single exponential, sech$(z)$, and \iso\
functions.  As described above, our $R$-band data were
corrected for dust at small $z$-heights by
using an extrapolation of the light profile from
intermediate $z$. Because of the clumpy nature of the dust in
UGC~7321, this procedure is uncertain at any given location to
$\sim\pm0.3$~mag~arcsec$^{-2}$. As with the $H$-band, fits
in the $R$-band were performed over the 
interval $-20''\le z\le +20''$.

The results of the $R$-band fits at $r$=0
are shown in Fig.~8. It can be seen that after
dust is accounted for, the $R$-band vertical light profile of UGC~7321
is reasonably reproduced over a large range in $z$
by a single exponential function of scale
height $h_{-z,c,R}\approx$\as{2}{84} on the $-z$ side of the profile and
$h_{+z,c,R}\approx$\as{3}{42} on the $+z$ side, for a mean of 
$\bar{h}_{z,c,R}\approx$152~pc. These values are slightly larger than derived
in the $H$-band, although the difference is only marginally significant 
to within expected errors (see
Sect.~5). Note however that there is 
possible evidence of some faint residuals to our
fits at
$\mu_{R}\ge$23.7~mag arcsec$^{-2}$ ($\sim$3\% of sky), particularly on the $-z$
side of the profile. Careful inspection of the $H$-band minor axis
profile (Fig.~3) shows that such residuals may also be marginally present in
these data. We return to this point in Sect.~6.1.3.

At larger galactocentric radii, we find that as with the $H$-band
data, the dust-corrected $R$ band vertical cuts cannot be
characterized over the full range of $z$ values by a single
exponential, sech$(z)$, or \iso\ function. In Fig.~9 we show the
$R$-band vertical profiles at $r=\pm$\am{0}{5} with the $H$-band
single-component sech$(z)$ fits from Fig.~6a 
overplotted. The residuals initially
seen in the $H$-band fits are now far more pronounced, occurring at
$R$-band surface brightness levels of \gsim22.8~mag~arcsec$^{-2}$, or
roughly 6\% of sky. Due to dust, at very small $z$-heights we are
unable to ascertain
how intrinsically peaked the $R$-band light profile is 
near the galaxy midplane. However, even taking
into account the uncertainties of the dust corrections, it is clear
that no one-component model can reproduce the data over the full
range of $z$.  Our $R$-band
fits thus give a second strong piece of evidence for the need for a  second
model component to fully reproduce the data. 

Fig.~10a \& b again show the observed 
$R$-band vertical brightness cuts at $r=\pm$\am{0}{5}, but this time
overplotted are scaled versions of the two-component \iso\
models derived from the $H$-band data at the corresponding 
radii. Both the scale
heights and the relative weights of the two components were taken to
be identical to those in the $H$-band fits at these radii. 
Only a simple scale factor
was applied to the total model profile to match the peak extrapolated
brightness of the $R$-band data. It can be seen that the
$H$-band models provide excellent fits to the $R$-band data, even at
large $z$ values. The small bump near  $z=12''$ in the upper panel is due to
an imperfectly subtracted foreground star.

Using two-component \iso\ models, fits were also performed to
the $R$-band brightness cuts at various other galactocentric
radii where the $H$-band profiles had already been fitted. The 
scale heights of the two components were
kept frozen, but the relative weights of the two components was
allowed to vary. These models continued to provide excellent fits to
the dust-corrected light profiles. 
As was the case for the $H$-band data, it was found
that it was necessary to increase the light contribution from the
larger scale height component as galactocentric radius increased,
and that a larger increase was required on the W side of the disk than
on the E side.

In spite of the higher signal-to-noise compared with the $H$-band data
at large $r$, near $|r|$\gsim\am{2}{0} we remain unable to formally
distinguish between one-component sech$(z)$ or \iso\ models 
and two-component \iso\ model fits, although
visually the two-component models appear to retain a slight advantage
(Fig.~11). Therefore we cannot rule out that at large galactocentric
radii the UGC~7321 is approximately isothermal (but see Section~6.1.1).

\section{Caveats, Uncertainties, the Uniqueness of the Fits}
Before we discuss the interpretation of the results inferred from
fitting the vertical light profiles of UGC~7321, we need to first assess the
caveats, uncertainties, and limitations of our fits.

Defining formal 
errors to the fit parameters we derived for the vertical light
structure of UGC~7321
is by no means straightforward (see also
Morrison \etal\ 1997; Pohlen \etal\ 2000). 
In all cases, 
uncertainties in our fits
are expected to arise from systematic effects, including sky
subtraction and flatfield errors, Poisson noise, and a host of
other factors, such as seeing, finite pixel size, scattered light, and 
light contamination from background
galaxies and field stars. However, it is nearly impossible to
formally quantify the magnitude of each of these sources of error,
hence we must attempt to make some estimate of their effects indirectly.

In the cases where we have fit our $H$-band data with a
single-component model, we initially fitted both the $+z$ and $-z$ sides of the
profiles independently. We found no systematic difference between the
scale heights derived for the two halves of the profile. Typically they
agreed to within a few per cent, with the maximum difference occurring
at $r$=0, where
the $h_{z}$ values derived for both the $+z$ and
$-z$ sides of the profile differ by $\sim$10\%. Although the $R$-band
data at $r=0$ show a similar result, implying this effect may be real, we
conservatively assume the reliability of any individual fit component
is $\sim\pm$10\%. The $R$-band data have higher signal-to-noise than
the $H$ band, but
suffer the additional uncertainty from the dust corrections, hence fit
errors are expected to be similar in the two bands.

Error assessment is even more complex in cases where we
have modelled our light profiles using two-component fits.
As discussed by Morrison \etal\ (1997), using
one-dimensional fits to galaxy light profiles it is impossible to
derive an entirely {\it unique} decomposition into a
multi-component model. Nonetheless, we find it is still possible to
derive some simple constraints on the characteristics of the vertical
light profile of UGC~7321 using a combination of the present data,
$B-R$ color maps (discussed further below), and a few simplifying assumptions.

It would undoubtedly be possible to produce adequate 
fits the the vertical light
profiles of UGC~7321 with more complex models comprised of additional disk
components. However, in the present analysis 
we adopt the simplest
models that can fully reproduce our data.
Even with a only a two-component model, it is still not possible to determine
a completely
unique combination of scale heights and relative weightings for the
two components to the final fit. However, one can put
reasonable limits on both quantities from the present data.

Let us consider the $H$-band light profiles extracted
at $r=\pm$\am{0}{5}, and the two-component \iso\ fits derived
independently on the $+z$ and $-z$ sides of the profiles,
for a total of four independent  fits.
For each of the four
fits, we find empirically that
the range of permitted scale heights for the two fit
components can be constrained within $\sim\pm$10\% by the requirement
that  the resulting total profile shows both a moderately
steep core as well as wings with a statistically significant amplitude at
$z$-heights $|z|\ge10''$. Likewise, from the fits to the four quadrants, 
we find differences of
\lsim10\% in the mean of the optimum values we derive.
Hence in this way, we have estimates of both the intrinsic uncertainties in our
fitting approach, as well as uncertainties arising from a range of 
systematic effects in the data. Thus for the remaining discussion, we
assume the uncertainties in the scale heights and relative weightings
in our two-component \iso\ model fits to be the sum
in quadrature of these two error terms, or $\sim$15\%.

A final caveat of our present fits is the inability to formally
distinguish between models comprised of two sech$(z)$ rather than
two \iso\ components.  
However, we chose to adopt the \iso\ function
(or isothermal disk)  due to the fact that it 
appears to be physically motivated in certain cases
(e.g., Dove \& Thronson 1993), and has been used to characterize the
old disk of the Milky Way (see Freeman 1991),
whereas the introduction of the sech$(z)$ model for galaxy disks
was largely based on empirical considerations alone
(cf. van der Kruit 1988). 
Moreover, although the exponential, sech$(z)$ and
\iso\ functions all exhibit very similar behavior at large $z$,
we find that using
an exponential for either of our two fit
components
 always produced a model light profile that was too strongly
peaked to reproduce our $H$-band observations at $|r|\le$\am{0}{25}.

\section{Interpretation of the UGC~7321 Fits}
\subsection{General Considerations}
As discussed above, our multi-component fits to the vertical light
profiles of UGC~7321 are not necessarily completely
unique.  What can we therefore say about their physical significance?
 
Firstly, it is worth noting that 
the requirement of a minimum of two disk components of
differing scale height to reproduce the light profile of UGC~7321 at
moderate galactocentric radii establishes that
{\it  even a seemingly simple galaxy disk system like
UGC~7321 has a complex structure and is not strictly isothermal}. 
However, greater physical 
insight can be gleaned by combining the results of our
fits with a high-quality 
$B-R$ color map of UGC~7321. 
In this way we find that our
measurements appear to readily lend themselves to a self-consistent physical
interpretation (see Section~6.1.1).

In analyzing the vertical structure of disk galaxies, a 
number of workers have reported either scale height changes with
radius, or else light excesses at large $z$ above those
predicted from a single exponential or isothermal disk
fit. These have been interpreted as either
evidence of a ``thick disk'' component (e.g., Burstein 1979;
van der Kruit \& Searle
1981a; Jensen \& Thuan 1982; van Dokkum \etal\ 1994;
Morrison \etal\ 1997; de Grijs \& Peletier 1997), or a possible ``halo''
component (e.g., Sackett \etal\ 1994; Abe \etal\ 1999). 
Although these interpretations
may well be justified for some galaxies, Dove \& Thronson (1993) showed
that an alternative in some cases 
could simply be a disk comprised of a sum of multiple
\iso\ components of differing scale heights, where the largest
$z$-height component is the oldest, similar to the early
suggestion of Oort (1932), to measurements of the Milky Way (e.g.,
Freeman 1991 and references therein), and similar to the models we
have proposed
for UGC~7321 (see also Kuijken 1991, who proposed an integral
representation of components). 

To some extent, the differences between
the ``thick disk'' and ``multiple component disk'' models is one of
semantics. However, the distinction is important, since true thick
disks analogous to that of the Milky Way (e.g., Gilmore \& Reid 1983) are
now frequently argued to result from external processes (e.g., Quinn \&
Goodman 1986; Morrison \etal\ 1998) and thus should not be
expected to be present in all disk galaxies. In contrast, a
multi-component disk may be a natural consequence of disk heating and
other internal evolutionary processes in virtually every galaxy. 
For a  given object,
optical color maps can serve an an important aid in distinguishing
between these different situations.

\subsubsection{Clues from Color Maps}
In an analysis of the edge-on galaxy IC~2531,
Wainscoat, Freeman, \& Hyland 
(1989) presented a vertical structural study
in which color maps were used as an aid to interpreting the
light profile
fits (see also Jensen \& Thuan 1982). 
In spite of the significant dust lane present in 
IC~2531, using optical color maps, Wainscoat \etal\  still were able to discern
the presence of a very thin,
blue, radially extended 
layer of stars along the galaxy midplane. Based on the existence
of this feature, they derived a fit to the
disk of IC~2531 using a model comprised of an ``old disk''
[a large scale-height 
exponential component  ($h_{z}$=533~pc for D=22~Mpc) contributing most of the
light], 
a dust layer, and finally, a ``young disk'' [a 
very small scale height exponential
component ($h_{z}=$67~pc) representing the 
thin layer of blue stars visible in the
color maps]. 

The $B-R$ color map of UGC~7321 is reproduced in
Fig.~12, with the location of our vertical light extractions
indicated. These data are described in detail in Paper~I. Here we
review some of the key trends.
As discussed in Paper~I, a small, red ($B-R\approx$1.5), 
nuclear feature is visible at
the center of UGC~7321, offset $\sim5''$ from the isophotal center of
the disk. Its nature is enigmatic (Paper~I).
On either side of this feature, very thin blue bands of
stars are visible along the disk midplane, sandwiched in a thicker,
much redder region, with $B-R\approx$1.2. 
This redder region extends to roughly $\pm20''$ on either
side of the disk center, and then shows a rather abrupt boundary. The
intersecting blue
layer of stars is seen to change in both color and morphology with radius; in
particular, near the
edge of the red ``sandwich'' region, this  
layer grows rather abruptly and
substantially thicker, achieving a roughly
constant thickness at radii beyond $\sim\pm30''$. Beyond that radius,
this region grows bluer with radius, reaching $B-R<$0.6 near the edge
of the disk. Finally,  out to at least $r\sim\pm$\am{2}{5}, the entire
disk of UGC~7321
is surrounded by a thicker but  flattened layer of
moderately red ($B-R\approx$1.1) stars. 

The complex color structure of the UGC~7321 disk is also evident along the
vertical direction. As discussed in Paper~I, at $|r|\le 5''$, vertical
color profiles show a reddening of a few tenths of a magnitude in
$B-R$ near
the galaxy midplane. However, at larger galactocentric radii a {\it
bluing} along the galaxy midplane becomes evident. This bluing becomes
more pronounced with increasing galactocentric radius, reaching
 $\Delta(B-R)\sim$0.45 near $r=$\am{2}{0}. In contrast, at 
larger $z$ heights ($|z|\ga 7''$), 
the color stays nearly constant at $B-R\sim$1.1 over the observed
range of galactocentric radii. Here we show that {\it the
 features apparent in the $B-R$ color map of UGC~7321 
show a number of  intriguing
correlations with the results of  our vertical structure fits.}

\subsubsection{Interpretation of the Models at Intermediate Galactic Radii}
At galactocentric distances $|r|\ge$\am{0}{5}, we argue that the two \iso\
components required to reproduce the light profile of UGC~7321
represent respectively the blue layer of  stars near the
midplane and the larger
$z$-height red component visible in our $B-R$ color map. 
That there is little apparent change in
thickness of these two components over the interval
\am{0}{5}$\le|r|\le$\am{2}{0} on our color map 
is consistent with the approximate constancy in
scale height of these two components with radius (to within $<$10\%) 
inferred from our
fits. Moreover, the $z$ ranges where each of the respective components
is seen to dominate the light distribution in our vertical fits in
Fig.~7 \& 10
corresponds very closely to
the boundaries of these components visible in Fig~12, or in plots of the 
vertical color profiles (see Fig.~15 of Paper~I). Although we noted in
Sections~4.5 \& 4.6 that we cannot readily formally distinguish
between one- and two-component models at $|r|>$\am{1}{5}, our color
map hints that a two-component disk may be present even at these radii.

As mentioned in Sect.~4, our model fits require an increase in the
total light contribution of the larger scale-height fit component as a
function of radius (see Fig.~7). 
At first glance this may seem surprising, given that
the disk of UGC~7321 gets increasingly {\it bluer} with radius. One
explanation could be that decreasing self-gravity and/or very low
surface densities in the outer disk
causes a gradual flaring of the star-forming layer, and hence a gradual
decrease of the concentration of light in the plane. Indeed, there appears to
be evidence of such behavior in {\it HST} images of UGC~7321 
(Gallagher \etal, in prep.). However, any subtle change
in scale height associated with this would be nearly impossible to infer from
our present data, given the decreasing signal-to-noise in the
outermost disk regions.

Based on  its blue color, it may also seem surprising that the disk component
associated with the layer of stars along the disk midplane
seems to be prominent even in the $H$-band data (see Sect.~4.6). 
Some of this effect likely stems from
the fact that the decreased flux of blue stars in the NIR is
partially offset by decreased internal absorption at longer
wavelengths.
Other possibilities are that due to limited
dynamical heating in the UGC~7321 disk, this layer
contains a mix of stellar populations, including some older, redder
stars, or that some NIR flux is emitted by M
supergiants (cf. Aoki \etal\ 1991). Indeed, high-resolution imaging 
data show evidence
for populations of young, luminous red supergiants or AGB stars
in UGC~7321 (Gallagher \etal, in
preparation).

As already noted in Paper~I, the presence of a redder, unresolved stellar
component at larger
$z$-heights in the UGC~7321 disk seems to give clear evidence that
{\it some dynamical evolution has occurred even in this thin 
galaxy}. A similar conclusion was reached by Bergvall \&
R\"onnback (1995) concerning the superthin galaxy ESO~146-014.
Furthermore, our simple double \iso\ model fits are consistent with the
suggestion of Dove \& Thronson (1993) that the different stellar
components of dynamically evolving 
galaxy disks can indeed be modelled as non-interacting,
quasi-independent, isothermal components.

\subsubsection{Interpretation of the Models of the Inner Disk Regions}
Assuming that the association we infer between our fitted disk
components and the vertical color structure of UGC~7321 holds at moderate
and large galactic radii, let us now reconsider the disk 
region
near $r$=0. In spite of the deceptively simple, single exponential fit
that we derived over the interval $-15''\le|r|$\lsim +$15''$, closer
inspection shows this to be the most
structurally complex portion of the UGC~7321 disk. 

As seen in Fig.~12, the region of UGC~7321
within $|r|\le$\as{20}{0} appears to be comprised of a minimum of three
distinct regions: (1) an old disk component like that seen at larger
$r$;  (2) an intermediate thickness redder component,
with a rather sharp boundary near $r=\pm 20''$; (3)
a very thin blue layer which grows slightly thicker over the interval
$0''<|r|<20''$, and thereafter appears to flare significantly, blending
with the blue disk region described above. Together
these components seem to offer a very natural 
 interpretation of the inferred exponential disk
structure of UGC~7321 near $r$=0---i.e., that the disk in this
region is  comprised of at least
three quasi-independent, isothermal or nearly isothermal 
components: a high $z$-height component, a moderate $z$-height component, and
a very small scale-height, marginally resolved
component along the disk midplane. 

Adopting such a model, we can
estimate the scale heights of each of these three components in a
manner similar to that used at larger galactocentric radii. 
The determinations were simplified by assuming the highest
$z$ height component has a scale height identical to that inferred for
the larger $z$-height component at $r\ge$\am{0}{5} ($\sim$423~pc).
For the other two components we then estimate $z_{0}\approx$\as{2}{4}
(116~pc) and $z_{0}\approx$\as{3}{9} (190~pc).
Note that the
color map in Fig.~12
shows the thin blue layer to be more strongly defined on the W side of the
galaxy than on the E side,
which explains why we inferred an exponential disk
structure near 
$r=-15''$, but why the light profile at $r=+15''$ appears to be
intermediate between a sech$(z)$ and an exponential at small $z$. Note
also that this model accounts for the residuals seen in the $R$-band
(and to a lesser degree the $H$-band) fits at $r=0$---i.e., they can
be interpreted as a continuation of the larger $z$-height disk
component seen at larger radii.

\subsection{Comparison of the Disk Structure of UGC~7321 with the
Milky Way and Other Galaxies}

\subsubsection{The Disk at Moderate Galactocentric Radii}

Interpretation of our UGC~7321 data in terms of multiple disk
components of differing scale heights (and hence differing
characteristic velocity dispersions) is certainly not a new idea. Such
components are known to exist in our own Milky Way (see Reid 1993;
van der Kruit 1999b for reviews) as
well as in other external galaxies (e.g., Jensen \& Thuan 1982;
Wainscoat \etal\ 1989;
van Dokkum \etal\ 1994). Moreover, the existence of disk components 
with different
stellar constituents and characteristic velocity dispersions 
should be predicted to arise in any galaxy disk via
dynamical evolutionary processes (e.g., Fuchs \&
Wielen 1987; Dove \& Thronson 1993; Just \etal\ 1996).

In the galaxy
IC~2531, Wainscoat \etal\ (1989) interpreted the small scale-height, blue 
layer of stars seen along the disk midplane
as a disk component analogous to the Milky
Way's ``young disk''. This layer is thus presumably
comprised primarily of OB stars, gas, and dust. Wainscoat \etal\
(1989) interpreted the underlying larger $z$-height disk component
as analogous to the Milky Way's ``old disk''. It is presumed to be
comprised of older,
redder, lower mass stars (see also Jensen \& Thuan
1982; van Dokkum
\etal\ 1994).
Indeed, based on their respective scale heights and colors, 
a similar interpretation seems natural for the
 disk components we
infer at intermediate radii in UGC~7321. Thus at intermediate
galactocentric radii, UGC~7321 can be said to
have an ``old disk'' with scale height $z_{0,O}\approx$423~pc, and a
``young disk'' with scale height $z_{0,Y}\approx$185~pc. 

In their analysis of IC~2531, Wainscoat
\etal\ used exponential rather than \iso\
fits\footnote{Note there is no
inconsistency here with our double \iso\ 
models of UGC~7321, since observations 
of K giants in the Milky Way have shown that its old disk
is indeed isothermal to at least 450~pc above the plane, while the apparent
overall exponential nature of the disk results from the presence of
additional dynamically colder stellar populations (Freeman 1993).}; 
thus recalling  
the relation
$z_{0}=2h_{z}$, we see that to within distance uncertainties, the 
scale height we measure for the young disk
of UGC~7321 is comparable to that of IC~2531 ($h_{z}\sim$67~pc) 
and of the Milky
Way ($h_{z}\sim$100~pc; Reid 1993; van der Kruit 1999b). 
However, the ``old disk'' of UGC~7321 is considerably
thinner compared with $h_{z}\approx$533~pc measured for the old disk of IC~2531 (Wainscoat
\etal\ 1989), $z_{0}\approx$1.1~kpc for NGC~6504 (van Dokkum \etal\
1994; see also de Grijs 1997 and references therein), 
and $h_{z}\sim$325~pc for
the Milky Way (Reid 1993).
Existing measurements thus seem 
consistent with the actively star-forming disks of most disk galaxies
having roughly similar scale heights, while the differences in their
global scale heights mainly stem from differences in the 
thicknesses of their older stellar components. 

The old disks of galaxies are expected to be formed from stars that have been
scattered to higher $z$-heights
via interactions with giant molecular clouds or scattering from spiral
arms (see Freeman 1991). Mild interactions
should also act to heat disks, at least temporarily 
(e.g., Reshetnikov \& Combes 1997). Hence it is not surprising that
compared to normal spirals,
the old disk should be significantly thinner in an isolated 
``superthin'' galaxy like UGC~7321.
One  obvious conclusion is then that  UGC~7321 has
undergone less dynamical heating and evolution than spirals like the Milky Way or
IC~2531.

Some galaxy formation models predict that  galaxies that are built very
slowly over time are expected to have very thin disks (e.g., Dove \& Thronson
1993; Noguchi 1998). While a long formation timescale may be a
required condition for the formation of a superthin galaxy,
our present results seem to indicate that 
the superthin appearance of UGC~7321 is
not solely a result of the conditions of formation, but also a  consequence of 
having avoided significant subsequent dynamical heating.  This lack
of heating
could be due to the low density of its disk, the lack of large
molecular complexes or spiral arms to scatter stars, the presence of 
an unusually massive dark
matter halo, and/or the consequence of having remained isolated during
most of its lifetime.

One last point worth noting is that the relative light contribution of
the young disk compared to the old disk that we infer for UGC~7321 is
considerably higher than that derived for galaxies such as IC~2531 or the
Milky Way. However, such a trend seems as expected for a blue,
gas-rich,  low surface brightness galaxy like
UGC~7321, where past star formation has likely been low, heating of
older populations has presumably been mild, and where 
the global disk colors are dominated by fairly young stars.

\subsubsection{The Inner Disk Regions}

As discussed in Sect.~6.1.3, at least two additional disk components 
appear to be present in
UGC~7321 at small galactocentric radii ($|r|\le20''$) in addition to
the old stellar disk: an intermediate scale-height component with
$z_{0}\sim$190~pc and a very thin component with $z_{0}\sim$116~pc. 

The bulk of the very small scale height component in the inner regions
of UGC~7321 ($|r|$\lsim15$''$)
has a blue color and 
gives rise the the observed globally exponential nature
of the vertical light profile near the galaxy center. This may be a
distinct disk region of UGC~7321, or it could be 
simply an extension of UGC~7321's 
star-forming disk from larger radii, but where the
scale height has become dramatically smaller due to the larger 
self-gravity near the disk
center.  As noted above, near the center of this region, a rather 
compact red feature
($\sim5''$ across) is seen in our color maps. This too may be a
distinct disk subcomponent, although it does not add any distinguishable
feature to the vertical light profiles. The extent of this
feature ($\sim$200~pc) 
is similar to that of  the CO-rich ``nuclear disk'' regions of the
Milky Way or NGC~891, where R$\sim$225~pc (Scoville \etal\ 1993). 
In the Milky Way, the
nuclear disk has a FWHM thickness of 
$\sim$65~pc (Scoville \etal\ 1993)---similar to the apparent 
extent of the feature in UGC~7321.
Although CO has recently been detected from the central regions of
UGC~7321 (Matthews \& Gao, in preparation), the resolution is insufficient
to map its spatial distribution.
We therefore collectively refer to this small scale-height red feature and the
surrounding small scale-height blue disk seen at $|r|$\lsim15$''$
as the ``young nuclear disk'' of UGC~7321.

Lastly, we comment on the red
disk region seen in UGC~7321 at $|r|$\lsim$20''$ (1~kpc). It was
shown in Paper~I that this region corresponds very closely in radial
extent to a light excess over a radial purely exponential brightness
profile (see also Fig.~2).
In a less inclined spiral, such a light excess would normally be
interpreted as evidence for a bulge.  However,
such an interpretation seems highly questionable for the case of
UGC~7321, which shows little central concentration of light in
broad-band images, and 
whose disk does not appear to ``bulge'' at all (e.g., Fig.~1). 
However, since
it is now believed that the bulges of late-type disk galaxies form
from inner disk instabilities (e.g., Combes \etal\ 1990; Norman, Hasan,
\& Sellwood 1996; Carollo 1999), one
possible explanation for  this
region is that it corresponds to some type of 
``proto-bulge''. Although this is
only one possible interpretation (see Paper~I),
we shall henceforth use this term to refer to this
disk component of UGC~7321.

\subsubsection{Summary of Disk Components in UGC~7321}
To summarize, in spite of its ``underevolved'' appearance,
UGC~7321 has a complex disk structure, 
with as many as 4 distinct components of
differing scale height (and hence of differing characteristic 
velocity dispersions) and differing stellar
make-up (as evidenced by their different $B-R$ colors and differing
radial extents). 
We term these: (1) the {\it old disk} with $z_{0,O}\approx$423~pc; (2) the
{\it young disk} with $z_{0,Y}\approx$185~pc; (3) the {\it young nuclear disk}
with $z_{0,N}\approx$116~pc; (4) the {\it proto-bulge} with
$z_{0,P}\approx$190~pc. The observed superposition of all the disk components
near $r=0$ yields an approximately exponential light distribution with a
{\it global} exponential scale height of
$h_{z,g}\approx$140~pc.  These results are consistent with a scenario
where the exponential nature of the vertical light profile of galaxy
disks is a result of dynamical evolution rather than purely the conditions of
its formation.

\section{Estimating the Vertical Velocity Dispersion in UGC~7321}
Measures of vertical velocity dispersions ($\sigma_{z}$) 
are of enormous interest in studies of disk galaxies, since they
can be used to gauge the vertical stability of disks (e.g., Fridman \&
Polyachenko 1984) and permit 
determinations of the stellar mass-to-light ratio
$\Upsilon_{\star}$, which in turn allows a determination of the total
mass contribution of the stellar disk. 
This is critical for interpreting galaxy rotation curves
in terms of mass models such that  the quantity and distribution of dark
matter can be inferred.

Unfortunately, stellar velocity dispersion measurements are quite
challenging observationally, hence they have been obtained only for a
handful of nearby disk galaxies (e.g., Bottema 1993 and references 
therein). For this reason, it is of interest to attempt to estimate
$\sigma_{z}$ indirectly, using measures of the disk scale height,
luminosity, and rotational velocity (e.g., Bottema 1993,1997; van der
Kruit 1999a). We now have the data to attempt this exercise for
UGC~7321. For simplicity, we limit ourselves to computing mean
velocity dispersions using the global disk scale heights from the
single-component fits
in Section~4.4.

For a disk with an exponential brightness profile, the vertical
velocity dispersion can be expressed as
\begin{equation}
\sigma_{z}(r)=\left[4\pi Gh_{z}\Sigma(r)\left(1-\frac{1}{2}e^{-|z|/h_{z}}\right)\right]^{1/2}
\end{equation}
\noindent where $\Sigma(r)$ is the mass surface density at a given radius
and $G$ is the gravitational constant (Wainscoat, Freeman, \& Hyland 1989).
For a sech$(z)$ vertical brightness distribution, one has instead
\begin{equation}
\sigma_{z}(r)=\left[1.7051\pi G\Sigma(r)h_{z}\right]^{1/2}.
\end{equation}
\noindent (e.g., van der Kruit 1999a).

One approach to estimating $\sigma_{z}$ involves assuming a reasonable
value for $\Upsilon_{\star}$ and converting the measured stellar
surface density to a mass density. However, this is 
particularly  uncertain for an
LSB galaxy like UGC~7321, which has strong radial and vertical color
gradients  and  an uncertain
metallicity. Instead we therefore adopt the approach of van der Kruit (1999a),
whereby the disk is assumed to be globally stabilized by a dark
halo. In that case, the semi-empirical criterion for global stability
of Efstathiou \etal\ (1982):
\begin{equation}
V_{rot}\left(\frac{h_{r}}{GM_{disk}}\right)^{1/2}\la 1.1.
\end{equation}
\noindent may be applied to estimate a total disk mass $M_{disk}$.

For UGC~7321, we have 
$V_{rot}\approx\frac{1}{2}(W_{20}-20)=106$~\kms (where we have
corrected the measured global \HI\ linewidth from Table~1 for 
turbulence; see Matthews, van Driel, \&
Gallagher 1998). 
Using the above relation, we derive
$M_{disk}\approx4.5\times10^{9}$\msun. With the additional approximation that 
the \HI\ follows the same
distribution as the stars, we can subtract the mass contribution for
the \HI, corrected for He, and arrive at
$\Sigma(0)\approx$110\msun~pc$^{-2}$. 
At $r=0$, taking $h_{z}$=140~pc and using Equation~7 
yields $\sigma_{z}(0)\sim$20.4~\kms. At $r$=\am{0}{5}, using a mean
$h_{z}$=145~pc and assuming $\Sigma(R)=\Sigma(0)e^{-r/h_{r}}$, from
Equation~8 we get
$\sigma_{z}($\am{0}{5}$)\sim$13.7~\kms, and at $r=1h_{r}$,
$\sigma_{z}(1h_{r})\sim$12.3~\kms. 
These values imply that UGC~7321 is a very
dynamically cold galaxy (cf. Bottema 1993). 
In spite of the approximations we have used, 
our $\sigma_{z}$ estimates show good
agreement with the {\it measured} 
values from Swaters (1999) for the moderately 
LSB dwarf spiral galaxy UGC~4325. He finds $\sigma_{z}(0)\sim19$~\kms and
$\sigma_{z}(1h_{r})\sim$13~\kms  (although note the measured $\sigma_{z}$
values are due to all mass components of the galaxy,
not just the stars). With $M_{B}=-$17.5 and $h_{r}=$1.6~kpc, 
UGC~4325 has a size and luminosity 
similar to UGC~7321, hence it is indeed expected that these two systems should
have similar stellar velocity dispersions (e.g., Bottema 1993).

\section{Summary}
We have presented measurements of the vertical structure of the low-luminosity
``superthin'' spiral galaxy UGC~7321. Using a combination of $H$- and
$R$-band data and $B-R$ color maps, we find that like more luminous
disk galaxies, UGC~7321 has a complex, multi-component disk structure,
indicating that some dynamical heating has occurred even in this thin
galaxy.
Our analysis has demonstrated the power of high-resolution
color maps as an aid in interpreting disk structural measures (see also
Jensen \& Thuan 1982; Wainscoat \etal\ 1989). 
Our findings for UGC~7321 also underscore the importance of utilizing
high resolution, multiwavelength data and sampling the
full range of $z$-heights of a galaxy disk when assessing
its vertical structure (see also Dove \& Thronson 1993). 

Near the disk center, the global $H$- and $R$-band vertical light
profiles of UGC~7321 can be reasonably characterized by a single exponential
function. We derive a global 
scale height at the disk center ($r=0$) of
$h_{z,g,H}\approx140\pm15$~pc in $H$-band and
$h_{z,g,R}\approx152\pm15$~pc in $R$-band. These are among the
smallest global vertical scale heights ever reported for any galaxy
disk, implying that UGC~7321 has not only a large disk axial ratio,
but also 
an intrinsically thin disk. Its
distance-independent ratio $h_{r}/h_{z}\sim$14 (at $r=0$) is also among the
largest reported for a galaxy disk. The overall
exponential appearance of the UGC~7321
disk at small galactocentric radii appears to result from the superposition of 
at least three quasi-independent  isothermal or nearly isothermal 
disk components of differing scale heights: 
(1) an ``old disk'' with $z_{0,O}\approx$423~pc; (2) a  ``young nuclear disk''
with $z_{0,N}\approx$116~pc; (3) a ``proto-bulge'' with
$z_{0,P}\approx$190~pc. Near the disk center, we estimate a mean
stellar velocity dispersion of $\sigma_{z}(0)\sim$20~\kms.

At roughly 20$''$ (1~kpc) from the galaxy center, the vertical light
profile of UGC~7321 becomes less peaked than an exponential near the 
midplane. At $|r|\ge20''$, although a sech$(z)$ function can provide a
rough global characterization of the data,
we find the best fits to the observed $H$-
and $R$-band light profiles are produced by  a linear combination of 
two isothermal components,
one representing
a continuation of the ``old disk'' with  $z_{0,O}\approx$423~pc, and
the other a
``young disk'' component 
with $z_{0,Y}\approx$185~pc. Both seem to contribute
roughly equal fractions to the total light in both the $H$ and $R$
bands, although the contribution of the old disk component increases
slightly as a function of galactocentric radius. Using this model 
we find no evidence for a
significant change in disk scale height as a function of radius  over
the interval $15''\le |r| \le$\am{2}{0}, 
although we cannot rule out changes of $<$10\%. 
From our present data it is unclear
whether
the ``young nuclear disk'' is a distinct entity, 
or whether it may be an extension of the young disk to small
galactocentric radii, where higher self-gravity results in a
dramatically smaller
scale height. At $|r|\ge$\am{2}{0}
our $B-R$ color map suggests a continuation of a multi-component disk
structure, although formal fits cannot rule out one-component models,
suggesting the possibility that the disk may become nearly isothermal
or sech$(z)$-like at large
galactocentric radii. If we use a single-component model to
characterize the disk at these radii, we find that near
$|r|$=\am{2}{0}
the ratio of disk
scale height to scale length becomes $h_{r}/h_{z}\sim$10, or roughly
one-third smaller than at $r$=0.

The mean scale height of the young disk of UGC~7321 is comparable to
that of the Milky Way and other external galaxies, suggesting that the
thickness of the star-forming layer in most spirals may be roughly similar.
The colors and $z$ distribution of the old disk component of UGC~7321
are consistent with it being comprised of an older population of
stars that have acquired higher velocity dispersions over time. 
However, the inferred scale height of UGC~7321's old disk  is
significantly lower than that of most spiral galaxies measured to date.  
Thus while some dynamical heating has
occurred in UGC~7321, it appears to have been extremely limited compared with
typical giant spirals. Results presented here and in Paper~I
seem to suggest that isolation, 
a low disk surface density, and a  massive dark matter halo
are key requirements for UGC~7321 to
 maintain its svelte appearance. Important additional tests of these
hypotheses will require stellar velocity dispersion measurements and
detailed mass modelling of galaxies like UGC~7321.

\acknowledgements
I am grateful for the financial support provided by a
Jansky Postdoctoral Fellowship from the National Radio Astronomy Observatory.
I acknowledge useful discussions
with J. S. Gallagher, L. S. Sparke, and B. Fuchs on various 
portions of this work, and I thank an anonymous referee for valuable 
comments. This publication made use of data products 
from the Two Micron All Sky Survey, which is a joint project of the 
University of
Massachusetts and the Infrared Processing and Analysis
Center/California 
Institute of Technology, funded by the National
Aeronautics and Space Administration and the National Science Foundation.

\newpage

\figcaption{$R$-band image of UGC~7321 obtained with the WIYN
telescope (see Paper~I for details). The
image is $\sim6'\times1'$. North is on top, west on the left.
Overplotted are pluses indicating the
location of the extracted vertical light profiles analyzed in this
paper. The larger symbol denotes the galaxy center ($r$=0).}

\figcaption{Radial surface brightness profiles in mag arcsec$^{-2}$ 
along the major axis of
UGC~7321: $H$-band (top); $R$-band (center); $B$-band (bottom). The
data were extracted along the disk major axis and folded about $r$=0, as
described in Section~3. Exponential fits to each of the profiles are
overplotted as dashed lines. The $H$-band data shown are not
photometrically calibrated, hence the absolute surface brightness
scale for these data
was estimated using the $H$-band data from the Two Micron All Sky
Survey (see Text), and was then offset by +1 mag arcsec$^{-2}$ for
display purposes.}

\figcaption{$H$-band minor axis light profile of UGC~7321. The absolute
photometric calibration was estimated using $H$-band data 
from the Two Micron All Sky Survey (see Text). Data points are shown
as plus signs. The best-fitting exponential model fits are overplotted as
dashed lines. Error bars were estimated as a sum in quadrature of Poisson,
flatfield, and sky subtraction errors.}

\figcaption{As in Fig.~3, but at $r=-15''$ (top) and $r=+15''$ (bottom).
In the lower panel a sech$(z)$ fit has also been plotted as a solid line.
Note in the lower panel that the exponential
fits are too peaked to reproduce the data at small $z$, while the
sech$(z)$ fit is somewhat too rounded at small $z$ and 
underestimates the profile wings at large $z$.}

\figcaption{As in Fig.~3, but at $r=-$\am{0}{5} (top) and
$r=+$\am{0}{5} (bottom). Here the
best-fitting exponential fits are shown as short dashed lines, the
best-fitting sech$(z)$ functions as solid lines, and the best \iso\
models as long dashed lines.}

\figcaption{$H$-band light profiles extracted parallel to the minor
axis of UGC~7321 at various galactocentric radii. The best-fitting
sech$(z)$ fits are overplotted. The panels depict light profiles
extracted at
the following locations: $(a)$: $r=-$\am{0}{5} (top);
$r=+$\am{0}{5} (bottom); $(b)$: $r=-$\am{1}{0} (top);
$r=+$\am{1}{0} (bottom); $(c)$: $r=-$\am{1}{5} (top);
$r=+$\am{1}{5} (bottom); $(d)$: $r=+$\am{2}{0}. Error bars were
estimated as in Fig.~3.}

\figcaption{As in Fig.~5, but with model fits comprised
of two \iso\ functions of scale heights $z_{0}=$\as{3}{8} and
$z_{0}=$\as{8}{7} respectively overplotted as long dashed lines. The
contribution of each of the two components to the total light is
indicated by short dashed lines.}

\figcaption{$R$-band minor axis profile of UGC~7321. The data are
shown as plus signs. The dotted line shows the data after
correction for dust absorption (see Text). The dashed lines show the
best-fitting exponential functions. Error were estimated using a
quadratic sum of Poisson, flatfield, and sky subtraction
errors and do not account for uncertainties in the correction for dust
absorption. Error bars are plotted only for every tenth data point.} 

\figcaption{$R$-band vertical light profiles extracted at $r=-$\am{0}{5}
(top) and $r=+$\am{0}{5} (bottom). The dotted lines show estimates of
dust-corrected profiles. The solid lines are the sech$(z)$ fits
derived from the $H$-band data at these respective radii. Error bars are plotted only for every tenth data point. }

\figcaption{As in Fig.~9, but with model fits comprised
of two \iso\ functions of scale heights $z_{0}=$\as{3}{8} and
$z_{0}=$\as{8}{7}, respectively, overplotted as long dashed lines. The
contribution of each of the two components to the total light is
indicated by a short dashed line. The bump near $z=12''$ in panel (a)
is due to an imperfectly-subtracted foreground star.} 

\figcaption{$R$-band vertical light profiles at $r=-$\am{2}{0} (top)
and $r=+$\am{2}{0} (bottom). The solid lines show the best sech$(z)$
fits. The long dashed line shows the best single-component \iso\
fit. The thick, medium dashed line shows a two-component \iso\
model with $z_{0,1}=$\as{3}{8} and
$z_{0,2}=$\as{8}{7}; the contributions of the two components are
overplotted as short dashed lines.}

\figcaption{Pseudocolor $B-R$ color map obtained from the 
WIYN imaging data of
Paper~I. Seeing was
$\sim$\as{0}{6}. Locations where the vertical light profiles 
were extracted and analyzed in the present paper
are marked by pluses. In this color scheme, 
the intrinsically
reddest regions of the disk
($B-R>$1.5) are seen as yellow. Other colors translate as
follows: reddish-orange 
($B-R\sim$1.2); greenish black ($B-R\sim$1.1); purple ($B-R\sim$1.0);
blue ($B-R\sim$0.8); black ($B-R<$0.7). The field of view shows
roughly the inner \am{3}{8} of the disk. $r=0$ is at the image center
and is indicated by the
symbol just to the left of the small yellow
feature.}

\end{document}